\documentclass[fleqn,usenatbib]{mnras}

\usepackage{newtxtext,newtxmath}

\usepackage[T1]{fontenc}

\DeclareRobustCommand{\VAN}[3]{#2}
\let\VANthebibliography\thebibliography
\def\thebibliography{\DeclareRobustCommand{\VAN}[3]{##3}\VANthebibliography}

\usepackage{graphicx}

\usepackage[T1]{fontenc}
\usepackage[utf8]{inputenc}
\usepackage{amsmath}
\usepackage{color}
\usepackage{mathtools}
\usepackage{hyperref}
\usepackage{natbib}

\let\caron\v

\newcommand{\dd}[1]{\mathrm{d}#1 \,}

\newcommand{\pdev}[2]{\frac{\partial #1}{\partial #2}}
\newcommand{\pdevn}[3]{\frac{\partial^{#3} #1}{\partial #2^{#3}}}

\newcommand{\dev}[2]{\frac{\mathrm{d}#1}{\mathrm{d}#2}}
\renewcommand{\v}[1]{\boldsymbol{#1}}

\newcommand{\crl}[1]{\langle #1 \rangle}

\title[CR transport in inhomogeneous media]{Cosmic-ray transport in inhomogeneous media}

\author[R. J. Ewart et al.]{Robert J. Ewart$^{1}$\thanks{E-mail: robertewart@princeton.edu}, Patrick Reichherzer$^{2,3}$, Shuzhe Ren$^{4}$, Stephen Majeski$^{1,5}$, Francesco Mori$^{2,6}$,\newauthor Michael L. Nastac$^{2,7}$, Archie F. A. Bott$^{2,8}$, Matthew W. Kunz$^{1,5}$, and Alexander A. Schekochihin$^{2,9}$
\\
$^{1}$Department of Astrophysical Sciences, Princeton University, Peyton Hall, Princeton, 08544, USA\\
$^{2}$Department of Physics, University of Oxford, Oxford, OX1 3PU, UK\\
$^{3}$Exeter College, Oxford, OX1 3DP, UK\\
$^{4}$St. Edmund Hall, Oxford, OX1 4AR, UK\\
$^{5}$Princeton Plasma Physics Laboratory, PO Box 451, Princeton, NJ 08543, USA\\
$^{6}$New College, Oxford, OX1 3BN, UK\\
$^{7}$St John's College, Oxford, OX1 3JP, UK\\
$^{8}$Trinity College, Oxford, OX1 3BI\\
$^{9}$Merton College, Oxford OX1 4JD, UK
}

\date{Accepted XXX. Received YYY; in original form ZZZ}

\pubyear{2025}

\begin{document}
\label{firstpage}
\pagerange{\pageref{firstpage}--\pageref{lastpage}}
\maketitle

\begin{abstract}
A theory of cosmic-ray transport in multi-phase diffusive media is developed, with the specific application to cases in which the cosmic-ray diffusion coefficient has large spatial fluctuations that may be inherently multi-scale. We demonstrate that the resulting transport of cosmic rays is diffusive in the long-time limit, with an average diffusion coefficient equal to the harmonic mean of the spatially varying diffusion coefficient. Thus, cosmic-ray transport is dominated by areas of low diffusion even if these areas occupy a relatively small, but not infinitesimal, fraction of the volume. On intermediate time scales, the cosmic rays experience transient effective sub-diffusion, as a result of low-diffusion regions interrupting long flights through high-diffusion regions. In the simplified case of a two-phase medium, we show that the extent and extremity of the sub-diffusivity of cosmic-ray transport is controlled by the spectral exponent of the distribution of patch sizes of each of the phases. We finally show that, despite strongly influencing the confinement times, the multi-phase medium is only capable of altering the energy dependence of cosmic-ray transport when there is a moderate (but not excessive) level of perpendicular diffusion across magnetic-field lines.
\end{abstract}




\section{Introduction}
\begin{figure*}
\centering
\includegraphics[width=17.78cm,height=7.62cm]{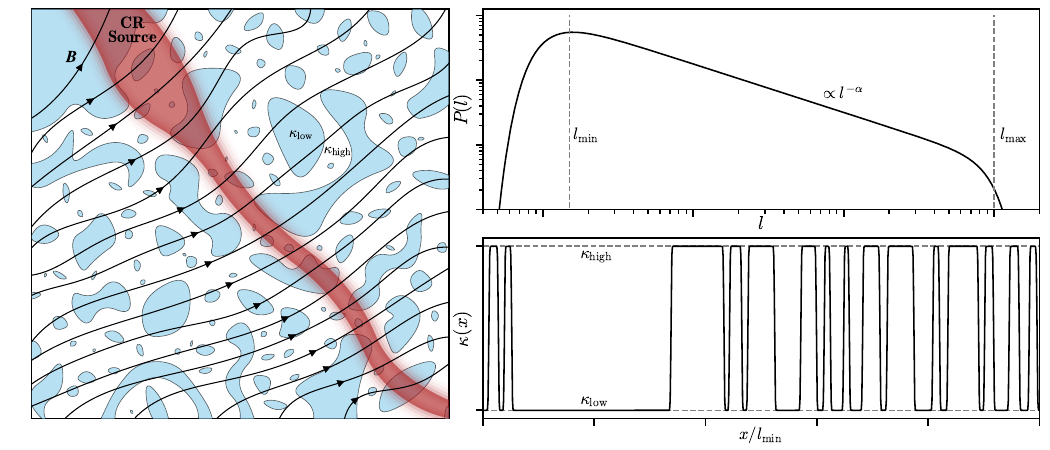}
\caption{\label{Fig:2}Cartoon of a patchy two-phase medium. In the left panel, the two-phase medium is depicted with regions of high parallel-diffusion coefficient in white and regions of low parallel-diffusion coefficient shaded in blue. CRs are sourced in localised areas of space, shown in red, across multiple field lines. The magnetic-field lines threading the media, to which CRs would be tied, are shown as black lines. The patchy stochastic structure is multi-scale, meaning that the distributions of patch lengths along magnetic-field lines, shown in the upper right panel, will typically be power laws. From these distributions, $P_{\mathrm{high}}(l)$ for high-diffusion patches and $P_{\mathrm{low}}(l)$ for low-diffusion patches, a single realisation of the spatially varying diffusion coefficient $\kappa(x)$ along a magnetic-field line can be constructed and is shown in the lower right panel.}
\end{figure*}
The transport of cosmic rays (CRs), particularly low-energy ({${\sim\! 1\!-\!100 \, \mathrm{GeV}}$}) CRs, has major implications for the launching of galactic winds (e.g., \citealt{Ruszkowski_2017,Quataert_2022b,Quataert_2022}), cosmological structure formation and galactic feedback (e.g., \citealt{Hopkins_2020,Quataert_2025}), re-ionisation in the early Universe (e.g., \citealt{Farcy_2025}), the morphology of the circumgalactic medium (e.g., \citealt{Giguere_2023,Weber_2025}), and much more besides (see, e.g., \citealt{Ruszkowski_2023} and references therein). In each of these situations, a common feature is that the CRs propagate not through a homogeneous medium with constant scattering rates but through a patchy, usually turbulent, environment that is frequently composed of multiple regions with distinct material properties. Indeed it has been convincingly proposed that an intermittent  picture of CR transport (\citealt{Kempski_2023,Lemoine_2023,Butsky_2024}) is required to resolve basic discrepancies between standard scattering models and CR observables (\citealt{Kempski_2022,Hopkins_2022}). The multi-phase nature of the medium may arise from macroscopically distinct fluid environments (e.g., \citealt{Sameer_2024,Beattie_2025,Hu_2025}) or microphysical differences caused by turbulence in the underlying fluid (e.g., \citealt{Squire_2021,Hopkins_2022b,Kempski_2023,Lemoine_2023,Reichherzera_2024}). The extreme diversity of such multi-phase environments naturally prompts two fundamental questions:
\begin{itemize}
\item For a given multi-phase medium, with each phase having a prescribed CR transport mechanism (ballistic streaming, diffusion, etc.), what are the average transport properties of the medium---in brief, what is the `mean-field' model of the medium?
\item What features of the multi-phase medium (e.g., filling fraction, topology, intermittency, etc.) impact the mean-field theory most?
\end{itemize}
In this paper, we answer these questions for a particularly simple model of CR transport. We consider the CRs to be tied to magnetic-field lines with a small diffusion coefficient perpendicular to the field, and assume that, while travelling along a single magnetic-field line, the CRs experience a spatially varying diffusion coefficient.\footnote{This is, effectively, a complementary set-up to that investigated by \cite{Butsky_2024}, in that both studies consider diffusion due to patchy substructure. We consider a population of CRs experiencing continuous diffusion over a large volume while confined to a~1D magnetic field, while \cite{Butsky_2024} considered CRs experiencing ballistic motion interrupted by events of extreme scattering that occupy a small fraction of the volume.} In a separate paper (\citealt{Reichherzerb_2025}), we consider the case in which this spatially varying diffusion coefficient arises from regions of fluid turbulence that are unstable to micro-scale plasma instabilities (\citealt{Schekochihin_2005,Squire_2023,Majeski_2024,Bott_2024}), which alter the local transport properties of CRs (\citealt{Ewart_2024,Reichherzera_2024}). In this paper, in pursuit of the general questions posed above, we are agnostic towards the microphysics of diffusion: the Universe has provided a multi-phase medium and it is the CRs' manifest destiny to propagate through it; the question is how they do so.

The central results, and outline of this paper, are as follows. In Section \ref{Section:1}, we first consider the mean-field theory. Naturally, on short length scales (equivalently, on short time scales), the transport is diffusive. This arises simply because CRs do not initially `know' that they are in a multi-phase medium---they experience only the diffusion coefficient characteristic of their local patch of space. If every CR initially knows only its local diffusion coefficient, then the effective diffusion coefficient is the spatially averaged diffusion coefficient of the system. On larger length scales (longer times) the transport is made more complicated by the fact that CRs will have sampled several different `patches' of the multi-phase medium. We show that, asymptotically, after the CRs have sampled many patches, their transport will again be diffusive but with an effective diffusion coefficient set by the harmonic mean of the local diffusion coefficient---an exact result (at late times), which we present in Section \ref{Section:1} (the proof is given in Appendix \ref{Section:A}). This means that the diffusive transport of CRs along magnetic-field lines is most sensitive to the areas having the smallest diffusion coefficients.
\begin{figure*}
\centering
\includegraphics[width=17.78cm,height=7.62cm]{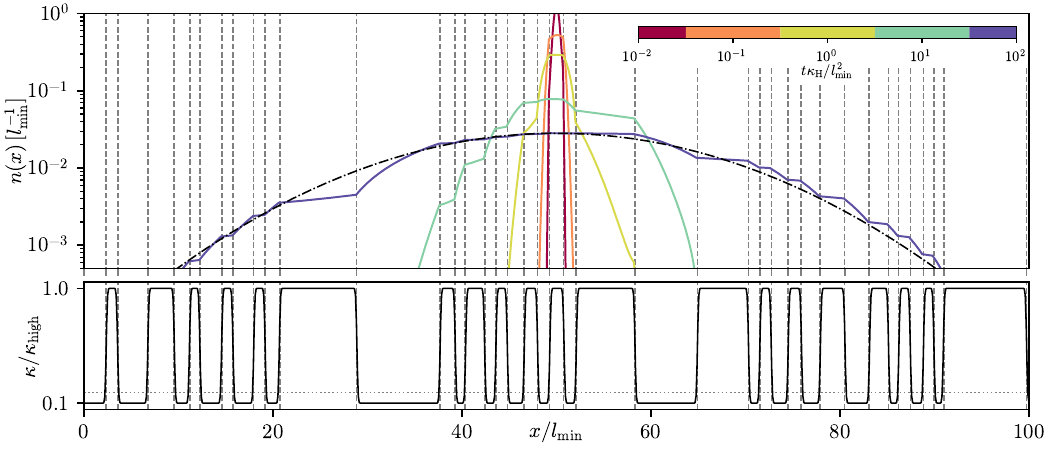}
\caption{\label{Fig:1}The solution of the diffusion model (\ref{eqn:1:1}) in a patchy environment. The upper panel shows the density of an initial point source in an inhomogeneous diffusive medium at five, logarithmically spaced, times. The spatially varying diffusion coefficient, shown in the lower panel, alternates between high and low values. The dashed lines demarcate the transitions between high- and low-diffusion regions, which create a characteristic `staircase' structure---the high-diffusion regions have flatter density profiles, separated by regions with larger gradients enabled in the low-diffusion regions. At late times, the overall density envelope spreads with a diffusion coefficient equal to the harmonic mean of the local diffusion coefficient (\ref{eqn:1:3})---this envelope is shown by the dash--dotted line. For this example, the distributions of patch sizes in high- and low-diffusion regions are $P_{\mathrm{high}}(l)= P_{\mathrm{low}}(l) \propto l^{-2.5}$.}
\end{figure*}
Thus, in a system where the local diffusion coefficient varies greatly, and over many length scales, the details of the transport naturally become sensitive to the nature of the spatial variation. In particular, since the CRs are diffusive at short times and diffusive again at large times, but with a smaller effective diffusion coefficient (since the harmonic mean is always smaller than the arithmetic mean), they must experience a transient period of sub-diffusion at intermediate times (cf. \citealt{Liang_2025}). This transient period is, naturally, a function of how long a CR can diffuse uninterrupted before fully sampling the multi-phase medium. The nature of the transport is therefore controlled by the sizes of the patches in the multi-phase medium, which are likely to be randomly distributed. As a model for such random multi-phase structure, which we refer to as `stochastic inhomogeneity', we consider the cartoon set-up shown in Figure \ref{Fig:2}: magnetic-field lines thread a two-phase multi-scale medium, and the patch lengths have a power-law distribution along a magnetic-field line.  In Section~\ref{Section:2}, we present a heuristic description of (and semi-analytic expression for) the resulting transient sub-diffusivity. We show that the predictions of this description are in robust agreement with numerical Monte-Carlo simulations of inhomogeneous diffusion equations. Despite the fact that the multi-scale, multi-phase nature of the medium can drastically alter the magnitude of the CR transport coefficient, we argue in Section~\ref{Section:3} (in the case of purely parallel diffusion) that the medium does not imprint its structure on the energy dependence of the CR transport (i.e., stochastic inhomogeneities affect the prefactor of the CR escape rate, but not its energy scaling). This is simply because the 1D geometry of parallel diffusion greatly constrains the path that CRs can take through the medium: they must tackle every low-diffusion patch head on. When a small amount of perpendicular diffusion is allowed, CRs instead navigate diffusively around the low-diffusion zones. If both the parallel and perpendicular diffusion coefficients are functions of the CR energy, then we should expect that the navigability of the medium will become a function of the CR energy. In Section~\ref{Section:3.1}, we present a heuristic description of the effect that this circumnavigation has on transport, arguing that, in the case of moderate (but not extreme) perpendicular diffusion, the stochastic inhomogeneity of the multi-phase medium can imprint itself on the energy dependence of the CR diffusivity. In Section~\ref{Section:4}, we summarise our progress and discuss future avenues of research.
\section{Theory}
\label{Section:1}
\begin{figure*}
\centering
\includegraphics[width=17.78cm,height=12.7cm]{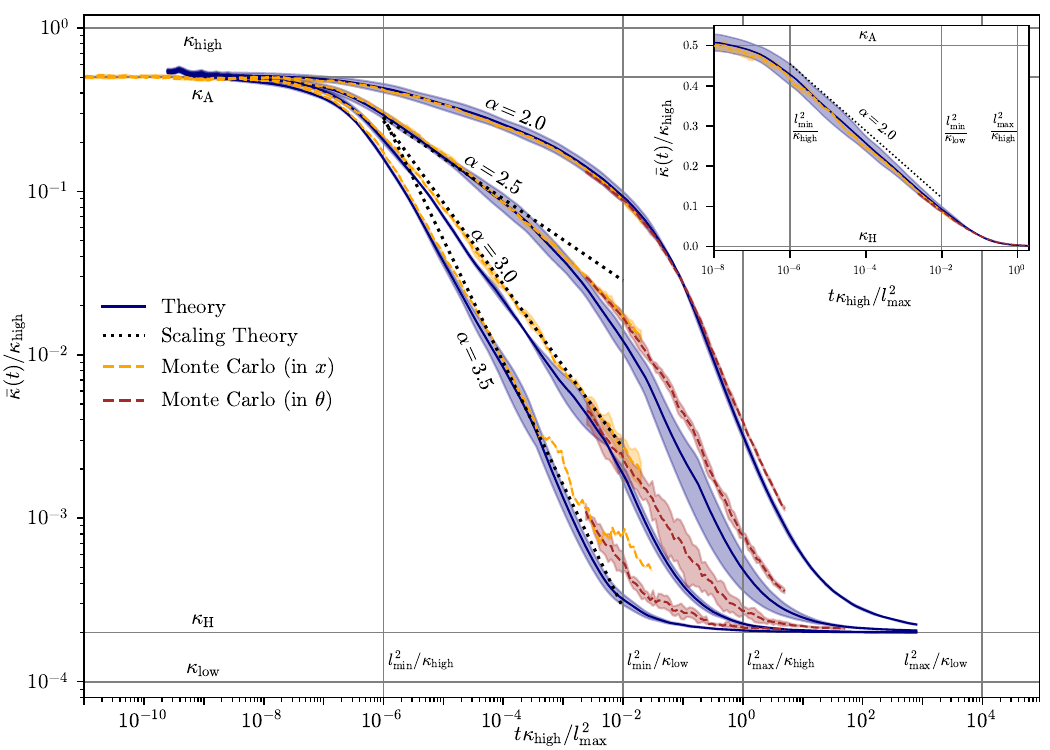}
\caption{\label{Fig:3}The running diffusion coefficient $\bar{\kappa}(t)$ of CRs propagating through inhomogeneous media with different $\alpha \equiv -\mathrm{d}\ln P_{\mathrm{high}}/\mathrm{d}\ln l$, where~$P_{\mathrm{high}}(l)$ is the probability of patch size $l \in [l_{\mathrm{min}},l_{\mathrm{max}}]$ for high-diffusion patches (but for these simulations, we chose $P_{\mathrm{low}} = P_{\mathrm{high}}$). Numerical Monte-Carlo solutions are shown (in yellow for random walks in position and red for random walks in pitch angle; see Appendix \ref{Section:C} for details). All diffusion coefficients begin at the arithmetic mean~(\ref{eqn:1:2p5}) and then reach the harmonic mean~(\ref{eqn:1:3}) after a period of sub-diffusion. The decay law of~$\bar{\kappa}(t)$ with time in the sub-diffusive regime can be computed (up to order-unity factors) easily from the scalings~(\ref{eqn:2:2}), shown as the black dotted lines. A more quantitative theoretical prediction, derived in Appendix~\ref{Section:B}, is shown by the the blue curves. The inset shows the~$\alpha = 2.0$ case on a semi-log scale, as the diffusion coefficient in this case is expected to decay logarithmically.}
\end{figure*}
We consider CRs propagating through a medium in which they experience differing local levels of scattering, and therefore differing levels of local spatial diffusion. This scenario can be conceptualised as the CRs propagating through distinct discrete patches as in Figure~\ref{Fig:2} (indeed this will later prove a very useful mental picture for understanding the transport). More generally, the density of CRs diffuses with an arbitrarily  spatially varying diffusion coefficient $\kappa(x)$:
\begin{equation}
\label{eqn:1:1}
\pdev{n}{t} = \pdev{}{x}\kappa(x)\pdev{n}{x},
\end{equation}
where $n$ is the CR number density (at a given CR energy) and~$x$ is a coordinate oriented along the magnetic-field line. It is this model---for now restricted to~1D (we lift this restriction in Section~\ref{Section:3.1})---describing transport only along magnetic-field lines that we wish to solve. 

By `solve', we mean that we wish to know the Green's function~$n(x,t;x_{0})$, i.e., the solution to (\ref{eqn:1:1}) that satisfies~{${n(x,0;x_{0}) = \delta(x-x_{0})}$}. Naturally, an exact such solution is a function of the particular realisation of the inhomogeneous environment~$\kappa(x)$. To the authors' knowledge, no useful analytical solution to (\ref{eqn:1:1}) exists that would be valid for all times. Instead, we will settle for statements about averaged properties, and asymptotics, of~$n(x,t;x_{0})$, which will be sufficient to develop a useful understanding of how CRs propagate through inhomogeneous environments. The most obvious such property would be a measure of the spread of CRs as a function of time, quantified by the `running diffusion coefficient'
\begin{equation}
\label{eqn:1:2}
\bar{\kappa}(t) = \frac{1}{2t}\left\langle\int \dd{x} (x-x_{0})^{2} n(x,t;x_{0})\right\rangle_{x_{0}},
\end{equation} 
where $\crl{...}_{x_{0}}$ denotes the average over all possible initial positions~$x_{0}$ of the CR. By writing $\bar{\kappa}(t)$ in terms of $n(x,t;x_{0})$ in this way, we have implicitly included the average over all the random displacements~$\Delta x(t)$ from the initial position $x_{0}$ of the individual CRs that comprise the density~$n(x,t;x_{0})$. Namely,~$\langle\langle \Delta x^{2}\rangle\rangle_{x_{0}}(t) = 2\bar{\kappa}(t)t$, where the double average is both over the CR trajectories and the initial positions $x_{0}$ where they originate---and, therefore, over the spatial structure of $\kappa(x)$. The physical meaning of~$\bar{\kappa}(t)$ is illustrated in Figure~\ref{Fig:1}, which shows the solution of the diffusion equation (\ref{eqn:1:1}) for a single realisation of a spatially inhomogeneous diffusion coefficient~$\kappa(x)$. At late times, manifestly, a converged value of~$\bar{\kappa}(t)$ is equivalent to the mean-field model for diffusion of CRs. However, as a proxy for the full solution~$n(x,t;x_{0})$, it is interesting to know~$\bar{\kappa}(t)$ for all times as a measure of how quickly a parcel of CRs diffuses away from a source (see Figure \ref{Fig:1}).

As stated above, at sufficiently early times, no CR will have diffused sufficiently far to experience variation from the diffusion coefficient~$\kappa(x_{0})$ of its initial position (i.e., it does not yet `realise' that it is in a multi-phase environment). Therefore, the average over all~$x_{0}$~is
\begin{equation}
\label{eqn:1:2p5}
\lim_{t \to 0}\bar{\kappa}(t)= \kappa_{\mathrm{A}} \equiv \lim_{X\to \infty} \frac{1}{X}\int_{0}^{X}\dd{X'}\kappa(X').
\end{equation}
We have arrived at the statement that, \textit{over short time intervals, the average CR diffusion coefficient is the arithmetic (`$\mathrm{A}$') spatial average of the varying diffusion coefficient} (this may be self-evident, but we prove it rigorously for completeness in Appendix \ref{Section:Z}). At long times, however, CRs will have had the opportunity to explore many different patches of the multi-phase medium, and it is not immediately clear how their transport should behave. To treat this behaviour correctly, we must separate the large length scale (over which we are interested in knowing the transport) from the small length scale (over which the spatial diffusion coefficient varies). This mathematical exercise is carried out formally in Appendix~\ref{Section:A}. The salient result is that the running diffusion coefficient (\ref{eqn:1:2}) of all CRs will converge at long times (large length scales) to the harmonic (`H') mean of the spatial diffusion coefficient:
\begin{equation}
\label{eqn:1:3}
\lim_{t\to \infty}\bar{\kappa}(t)= \kappa_{\mathrm{H}} \equiv \lim_{X \to \infty}\frac{X}{\displaystyle\int_{0}^{X}\!\!\frac{\mathrm{d}X'}{\kappa(X')}}.
\end{equation}
The mnemonic here is that \textit{diffusion coefficients add like capacitors in series}.\footnote{This analogy is precise (i.e., there is a mapping of the diffusive problem~(\ref{eqn:1:1}), appropriately discretised, onto a system of resistors and capacitors; see, e.g., \citealt{Bouchaud_1990}). The harmonic mean therefore emerges as a natural way of averaging in a variety of contexts. For diffusive processes, it is often used to describe transport of particles on lattice sites or through disordered media (e.g., \citealt{Alexander_1981,Bouchaud_1990,Igloi_1999,Miyaguchi_2016,Schlaich_2025}) or as the natural way to average opacities in radiative-transfer problems (e.g., \citealt{Rybicki_1979}).} This is an exact result for CRs diffusing perfectly along magnetic-field lines, but can be inferred physically either by positing that it is the CR scattering rates~($\propto \kappa^{-1}$) that should be averaged arithmetically (cf. \citealt{Butsky_2024}) or by noting that CRs are reflected many times off low-diffusion regions before passing through them, giving low-diffusion regions a stronger weight (cf. \citealt{Kumiko_2008,Reichherzera_2024}).

Thus, we have arrived at an answer to the question of what effective diffusivity one needs to describe CR transport in a multi-phase medium: the arithmetic mean~$\kappa_{\mathrm{A}}$ at early times and the harmonic mean~$\kappa_{\mathrm{H}}$ at late times. In a two-phase medium (i.e., one where the diffusion coefficient can take only two values,~$\kappa_{\mathrm{high}}$ and~$\kappa_{\mathrm{low}}$---see the cartoon in Figure~\ref{Fig:2}), equations~(\ref{eqn:1:2p5}) and~(\ref{eqn:1:3}) collapse to depend on a single parameter: the linear (along the magnetic-field line) filling fraction~$f$ of the low-diffusion phase, viz.
\begin{equation}
\label{eqn:1:4}
\begin{split}
\kappa_{\mathrm{A}} &= f\kappa_{\mathrm{low}} + (1-f)\kappa_{\mathrm{high}}\approx (1-f)\kappa_{\mathrm{high}}, \\  \kappa_{\mathrm{H}} &= \frac{1}{\displaystyle\frac{f}{\kappa_{\mathrm{low}}} + \frac{1-f}{\kappa_{\mathrm{high}}}} \approx \frac{\kappa_{\mathrm{low}}}{f}.
\end{split}
\end{equation}
It may seem as though this is immediately a fantastic economy: we do not need to know the detailed intricacies of the multi-phase medium, but only the effective diffusion coefficient for each phase and the corresponding filling fraction~$f$. All other details can be left to turbulence theorists. There are, however, some important cases where we too will wish for more detail. First, the CR diffusion coefficient must find a way to transition from its initial high value~$\kappa_{\mathrm{A}}$ to its late-time low value~$\kappa_{\mathrm{H}}$: the CRs must undergo a period of transient sub-diffusion\footnote{Via the Cauchy--Schwarz inequality, it is easy to show that the arithmetic mean~$\kappa_{\mathrm{A}}$ will always be greater than the harmonic mean~$\kappa_{\mathrm{H}}$ (with equality achieved only for spatially homogeneous diffusion coefficients); so the transport at intermediate times will always be sub-diffusive rather than super-diffusive.}. Secondly, the scattering frequency is not only a function of the position of the CR, but also of its energy, which means that the stochastically inhomogeneous medium appears different to CRs of different energies. Both of these features---the transient sub-diffusion and the energy dependence of transport---depend on the finer details of the multi-phase medium.

\section{Transient Sub-diffusion}
\label{Section:2}
So far we have stated only vaguely that the transient sub-diffusive behaviour lasts until `CRs have sampled multiple phases'. In the case where the multi-phase medium is also multi-scale, this can take a long time (as the time taken to cross a length $l$ diffusively scales as $l^{2}$), implying that the sub-diffusive transient may be relevant for understanding CR residence times. In this section, we explore this possibility in detail. 
\subsection{Flights through initial patches}
Consider first a simplified but instructive example of a two-phase medium with extremely disparate diffusion coefficients,~${\kappa_{\mathrm{low}}\ll \kappa_{\mathrm{high}}}$ (such media have been proposed by, e.g., \citealt{Reichherzera_2024}). To capture the multi-scale nature of such a set-up, we need two further pieces of information: \textit{the probability distribution functions (pdfs) of the sizes of the low- and high-diffusion patches along magnetic-field lines},~$P_{\mathrm{low}}(l)$ and~$P_{\mathrm{high}}(l)$. A schematic of this is shown in Figure \ref{Fig:2}: from a given patchy substructure, one extracts a magnetic-field line and measures the lengths of patches along it, creating a histogram of the possible lengths\footnote{The `filling fraction'~$f$ used in, e.g.,~(\ref{eqn:1:4}) is the filling fraction along magnetic-field lines as opposed to the volume-filling fraction. In \cite{Reichherzerb_2025}, we find that these two quantities are comparable for the patches of the fields that are unstable to a specific plasma micro-instability, but it is not inconceivable that they could be distinct for different forms of turbulence.}. Now consider the motion of CRs, uniformly distributed in this set-up. Initially the displacement of CRs sourced within the high-diffusion patches scales as~$\crl{\Delta x^{2}} \sim \kappa_{\mathrm{high}}t$. In the low-diffusion patches, however, the CRs are diffusing so slowly that they are practically frozen and do not contribute strongly to the global diffusion (yet). This means that the average diffusion coefficient is approximately~$\kappa_{\mathrm{high}}$ multiplied by the fraction~$1-f$ of the CRs that reside in the high-diffusion region, in line with the exact result~(\ref{eqn:1:4}). However, once these CRs (those sourced in high-diffusion patches) have diffused to a scale comparable to the size of their initial patch, they encounter a low-diffusion patch and thus are naturally confined to that scale, only diffusing farther on a much longer time scale. Therefore, after a time~$t$, only those CRs that were sourced in high-diffusion patches larger {than~${\sim \sqrt{\kappa_{\mathrm{high}}t}}$} still contribute a diffusion coefficient~$\kappa_{\mathrm{high}}$. Therefore, the running diffusion coefficient will scale roughly with the fraction of CRs that are sourced in high-diffusion patches that are larger than $\sqrt{\kappa_{\mathrm{high}}t}$. If we assume CRs are sourced in high- and low-diffusion patches in a way that is statistically uniform (but not necessarily spatially uniform---see Figure \ref{Fig:2}) then this gives
\begin{equation}
\label{eqn:2:1}
\bar{\kappa}(t) \sim \kappa_{\mathrm{high}}\displaystyle\frac{\displaystyle\int_{\sqrt{\kappa_{\mathrm{high}}t}}^{l_{\mathrm{max}}}\dd{l}P_{\mathrm{high}}(l)l}{\displaystyle\int_{l_{\mathrm{min}}}^{l_{\mathrm{max}}}\dd{l} \left[P_{\mathrm{high}}(l)+ P_{\mathrm{low}}(l)\right]l}.
\end{equation} 

Consider the patch sizes to be multi-scale with a pdf of patch sizes, as in Figure \ref{Fig:1}, that scales for both high- and low-diffusion patches like~$P(l) \sim l^{-\alpha}$ between~$l_{\mathrm{min}}$ (the smallest patch) and~$l_{\mathrm{max}}$ (the largest patch). Then (\ref{eqn:2:1}) becomes
\begin{equation}
\label{eqn:2:2}
\bar{\kappa}(t) \sim (1-f)\kappa_{\mathrm{high}}\begin{cases} 1 - \left(\displaystyle\frac{\kappa_{\mathrm{high}}t}{l_{\mathrm{max}}^{2}} \right)^{(2-\alpha)/2} \,\, &\text{for} \,\,\, \alpha < 2, \\[15pt] \left.\ln\left(\displaystyle\frac{l_{\mathrm{max}}^{2}}{t \kappa_{\mathrm{high}}} \right)\right/ \ln\left(\displaystyle\frac{l_{\mathrm{max}}^{2}}{l_{\mathrm{min}}^{2}} \right) \,\, &\text{for} \,\,\, \alpha = 2, \\[15pt] \left(\displaystyle\frac{t \kappa_{\mathrm{high}}}{l_{\mathrm{min}}^{2}} \right)^{(2-\alpha)/2}\,\, &\text{for} \,\,\, 4 \geq \alpha > 2, \\[15pt]
\left(\displaystyle\frac{t \kappa_{\mathrm{high}}}{l_{\mathrm{min}}^{2}} \right)^{-1}\,\,\, &\text{for} \,\, \alpha > 4. 
\end{cases}
\end{equation}
The anomalous scaling~$\bar{\kappa}(t) \sim  l_{\mathrm{min}}^{2}/t$ for~$\alpha > 4$  occurs because the large patches are so rare in this case that the multi-phase medium is effectively single scale: the running diffusion coefficient is dominated by those CRs that are confined to their original patch of size~$l_{\mathrm{min}}$, giving~$\bar{\kappa}(t) \sim l_{\mathrm{min}}^{2}/t$. The same~$t^{-1}$ scaling arises after a time~${t \sim l_{\mathrm{max}}^{2}/\kappa_{\mathrm{high}}}$, by which point all CRs have typically left their initial patch. 

We stress that this picture of sub-diffusion is simply that of book-keeping the number of CRs that are still participating in their high-diffusion flight through their initial patch (not dissimilar to `first-passage processes'; see \citealt{Redner_2001}). We verify these predictions numerically in Figure~\ref{Fig:3}, showing good agreement with our scalings for the early- and late-time diffusion coefficients~(\ref{eqn:2:2}) as well as the dependence of the sub-diffusive transient on the patch-size distribution. While the scaling theory that we have proposed captures the physics of the sub-diffusive transient, in the spirit of due diligence, it is possible to construct a quantitative prediction for the full time dependence of~$\bar{\kappa}(t)$ for arbitrary (not just two-phase) $\kappa(x)$. We do this  in Appendix \ref{Section:B} (essentially combining the exact results~(\ref{eqn:1:2}) and~(\ref{eqn:1:3}) with the scaling theory~(\ref{eqn:2:2})). The results are in good agreement with the numerical simulations shown in Figure~\ref{Fig:3}.

\subsection{Energy dependence of CR transport}
\label{Section:3}
\begin{figure}
\centering
\includegraphics[width=8.68cm,height=8.68cm]{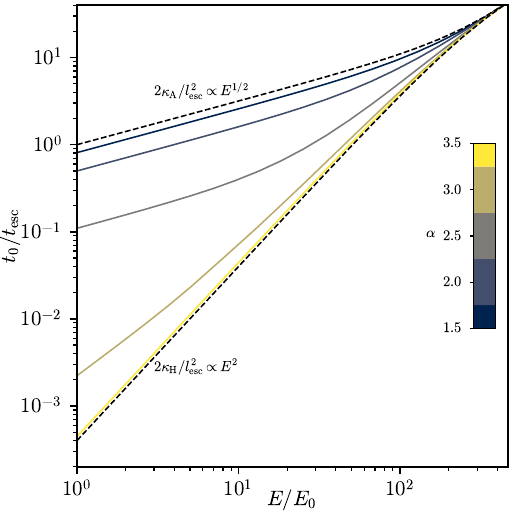}
\caption{\label{Fig:4}The escape rate (inverse time for a CR to diffuse past a distance~$L_{\mathrm{esc}}$, normalised to $t_0 = l_{\mathrm{esc}}^{2}/2\kappa_{\mathrm{A}}(E_{0})$) as a function of the CR energy~$E$ relative to a reference energy $E_{0}$. In this numerical set-up (chosen entirely for illustrative purpose), we considered the CRs to have an energy-dependent diffusion coefficient~$\propto E^{1/2}$ in the low-diffusion regions and~$\propto E^{2}$ in the high-diffusion regions (cf. \citealt{Reichherzera_2024}). We chose~{${l_{\mathrm{esc}} = 10\,l_{\mathrm{min}} = l_{\mathrm{max}}/10}$}. As a result, for low~$\alpha$ (where large patches are frequent), the escape rate is dominated by CRs' diffusion in long high-diffusion patches, while for~$\alpha > 2$ long patches become rarer, so slow diffusion dominates the CR escape rate.}
\end{figure}
\begin{figure*}
\centering
\includegraphics[width=17.78cm,height=10.67cm]{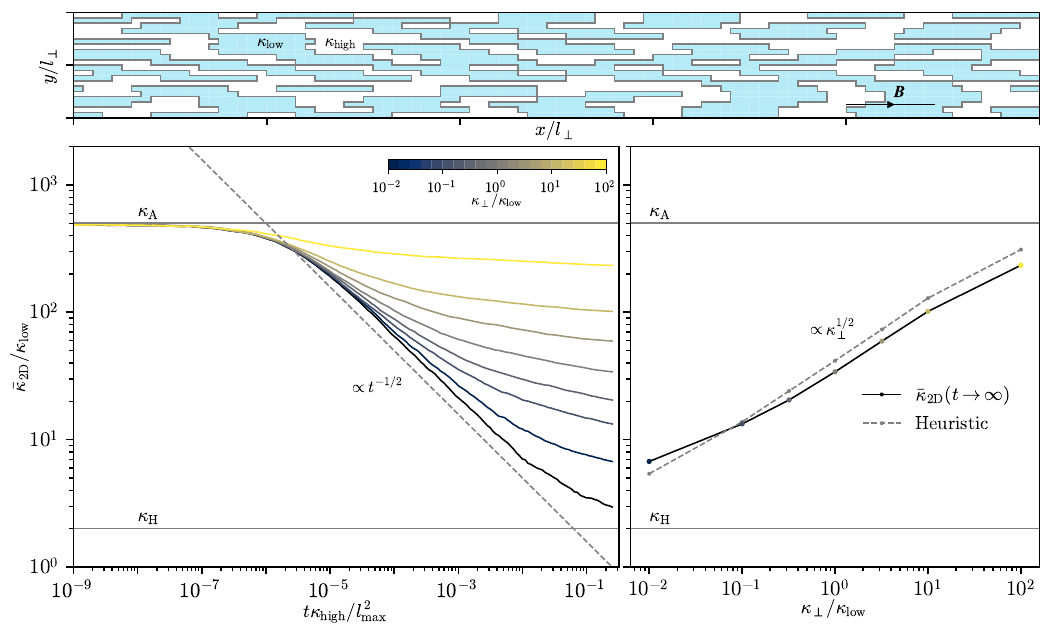}
\caption{\label{Fig:C1}Effective diffusion coefficient in a 2D diffusive medium. The upper panel shows a small sample of our model diffusive medium through which CRs propagate. The lower-left panel shows the running parallel diffusion coefficient $\bar{\kappa}_{\mathrm{2D}}(t)$; the black line is the case with~$\kappa_{\perp}=0$ (the 1D running diffusion coefficient~$\bar{\kappa}(t)$). The lower-right panel shows the converged effective parallel diffusion $\bar{\kappa}_{\mathrm{2D}}(t\to \infty)$, which interpolates between the harmonic mean~$\kappa_{\mathrm{H}}$ (CRs must pass through every patch on the field line) and the arithmetic mean~$\kappa_{\mathrm{A}}$ (CRs diffuse so rapidly perpendicularly that they uniformly sample the diffusion coefficient). This curve is in reasonable agreement with the simple heuristic (grey line) given by (\ref{eqn:4:4}) from (\ref{eqn:4:3p5}). For this simulation, $\kappa_{\mathrm{high}}/\kappa_{\mathrm{low}} = 10^{3}$, the pdf of high-diffusion patches was chosen to be $P_{\mathrm{high}}(l) \propto l^{-3}$ with $l_{\mathrm{max}}/l_{\mathrm{low}} = 400$, whereas the low-diffusion patches were chosen to be single-scale:~$P_{\mathrm{low}}(l) = \delta(l-l_{0})$ with~$l_{0}=2l_{\mathrm{min}}$ (giving~$f=0.5$). In the perpendicular direction, the system had a single correlation length~$l_{\perp} = l_{\mathrm{min}}/5$.}
\end{figure*}
Having understood the transport of CRs with a given (variable) diffusion coefficient, let us consider the signature of a multi-phase medium on CR observables. In particular, to compare CR transport in a multi-phase medium to the classic `leaky-box' model, one must compute the CR escape rate~$t_{\mathrm{esc}}^{-1}$ beyond a certain distance~$L_{\mathrm{esc}}$: the average rate at which CRs first pass a distance $L_{\mathrm{esc}}$ from their initial location. This is a subtly different problem to computing the diffusion coefficient~$\bar{\kappa}(t)$, as the escape rate is an average at fixed length rather than fixed time, but the central concepts are easily translated. As can be seen from Figure~\ref{Fig:3} (or from~(\ref{eqn:1:3}) and~(\ref{eqn:1:4})), the escape rate at short length scales~$L_{\mathrm{esc}} \ll l_{\mathrm{min}}$ is set by the rate at which CRs can diffuse through the high-diffusion regions ($t_{\mathrm{esc}}^{-1}\sim \kappa_{\mathrm{A}}/L_{\mathrm{esc}}^{2}$), while at the escape rate at long length scales~{${L_{\mathrm{esc}} \gg l_{\mathrm{max}}}$}, is dominated by the rate at which CRs can diffuse through the low-diffusion medium~($t_{\mathrm{esc}}^{-1}\sim \kappa_{\mathrm{H}}/L_{\mathrm{esc}}^{2}$). 

On intermediate scales~$l_{\mathrm{min}} \ll L_{\mathrm{esc}} \ll l_{\mathrm{max}}$, there is a competition between the majority of CRs diffusing with a diffusion coefficient~$\kappa \sim \kappa_{\mathrm{H}}\sim \kappa_{\mathrm{low}}$ and a select few that escape much quicker with~$\kappa\sim\kappa_{\mathrm{high}}$. The fraction of CRs that escape with rate $\kappa_{\mathrm{high}}/L_{\mathrm{esc}}^{2}$ is, by the same token as in the argument for~(\ref{eqn:2:1}), the fraction that were lucky enough to be born in high-diffusion patches longer that $L_{\mathrm{esc}}$, which we denote~$f_{l>L_{\mathrm{esc}}}$. This quantity is simply a property of the stochastically inhomogeneous diffusion coefficient and can be computed in the same manner as (\ref{eqn:2:1}) and (\ref{eqn:2:2}) for different $P(l)$. Without doing this explicitly, one may still write down a rough estimate of the mean CR escape rate:
\begin{equation}
t_{\mathrm{esc}}^{-1} \sim \frac{\kappa_{\mathrm{high}}(E)}{L_{\mathrm{esc}}^{2}}f_{l>L_{\mathrm{esc}}} + \frac{\kappa_{\mathrm{low}}(E)}{L_{\mathrm{esc}}^{2}}(1- f_{l>L_{\mathrm{esc}}}).
\end{equation}
The salient property of this mean escape rate is that all information about the stochastic inhomogeneity of $\kappa(x)$ is encoded in the $L_{\mathrm{esc}}$-dependence, \textit{not the energy dependence}. For a specified energy dependence of the diffusion coefficients in the two phases, this can lead to a break scale (as shown in Figure~\ref{Fig:4}) since there are combinations of energies and length scales for which~${\kappa_{\mathrm{high}}(E)f_{l>L_{\mathrm{esc}}} < \kappa_{\mathrm{low}}(E)(1-f_{l>L_{\mathrm{esc}}})}$ despite the large disparity in the diffusion coefficients,~$\kappa_{\mathrm{high}}\gg \kappa_{\mathrm{low}}$. Crucially, however, the energy dependence of the escape rate on either side of this break scale will be determined by the energy dependence of the single phase, regardless of the distribution of patch sizes: an observer could not determine the nature of the inhomogeneous medium solely from a measurement of the escape rate. In principle, it is possible, with much fine tuning (cf. \citealt{Kempski_2022}), to imagine a multi-phase medium composed of a very large number of phases each with a different energy-dependent diffusion coefficient. In such a system\footnote{This may, however, be the correct way to map (somewhat unnaturally) our model of a multi-phase medium onto one of ballistic transport mediated by strong scattering off structures of a particular scale (as in, e.g., \citealt{Kempski_2023,Butsky_2024}).}, each CR of a different energy would effectively see an entirely different multi-phase medium and a measurement of the energy dependence of CR transport would provide (some) information about the underlying structure of the ambient medium. Within our model diffusive problem, a much more natural way in which CRs of different energies could experience different environments would be via perpendicular diffusion. We study this possibility in the next section.
\section{The role of cross-field diffusion}
\label{Section:3.1}
So far we have considered parallel diffusion along magnetic-field lines assuming no perpendicular diffusion. While the presence of a multi-phase medium can naturally create interesting transient features shown in Figures~\ref{Fig:3} and~\ref{Fig:4}, we showed in Section \ref{Section:3} that the diffusion, and in particular its energy dependence, becomes independent of patch-size distributions at long times and large length scales. The reason for this simplicity is that the underlying geometry is simple: to get from A to B, all CRs must travel through the same intervening medium. While this makes it possible to formulate the exact laws (\ref{eqn:1:2}) and (\ref{eqn:1:3}), it also ignores the possibility of more subtle CRs trajectories, e.g., cases where CRs `dodge' the low-diffusion patches. On physical grounds, it is also clear that transport off of field lines (e.g., \citealt{Desiati_2014,Kempski_2023,Lemoine_2023}), or the wandering of field lines themselves (e.g., \citealt{Enslin_2003,Lazarian_2014}) will play a significant role in CR diffusion. Therefore, in this section, we address the role of perpendicular diffusion. 

The most obvious question to address is `when is perpendicular diffusion unimportant?'. The framework that we have presented requires the CRs to be able to sample very many structures along a magnetic-field line. Clearly, for perpendicular diffusion to be unimportant, the CR must be able to pass through a structure in the parallel direction before diffusing through it perpendicularly. The largest structure has some parallel scale $l_{\mathrm{max}}$ along the magnetic field, and the time that it would take a CR to diffuse through such a structure, with parallel-diffusion coefficient $\kappa_{\mathrm{low}}$, is $\sim l_{\mathrm{max}}^{2}/\kappa_{\mathrm{low}}$. If the CR has a perpendicular-diffusion coefficient~$\kappa_{\perp}$ in the low-diffusion patch and this patch has a minimum perpendicular scale~$l_{\perp}$, then the condition that the typical CR crosses the full parallel length of the patch before crossing its perpendicular length is 
\begin{equation}
\label{eqn:3:1}
\kappa_{\perp} \lesssim \frac{l_{\perp}^{2}}{l_{\mathrm{max}}^{2}}\kappa_{\mathrm{low}}.
\end{equation}
While it is then still true that CRs will eventually traverse the perpendicular scale of a structure, they will have sampled very many structures in the parallel direction before such an event occurs, by which time, from (\ref{eqn:1:4}), their parallel-diffusion coefficient will be well converged to $\sim\kappa_{\mathrm{low}}/f$. Thus, if (\ref{eqn:3:1}) holds, perpendicular diffusion is irrelevant and CRs can be considered basically to follow magnetic-field lines, slowly.

When (\ref{eqn:3:1}) is not satisfied, $(\kappa_{\perp}/\kappa_{\mathrm{low}})^{1/2}$ sets the aspect ratio of the patches that CRs will preferentially diffuse out of in the perpendicular direction as opposed to diffusing through in the parallel direction. This naturally sets the question of percolation: many routes through the medium are available to CRs, some of which are favoured or disfavoured due to higher or lower diffusion structures around which the CRs must navigate (cf. geometrical optics; \citealt{Meerson_2019}). The problem is hard (but also rich; see, e.g., \citealt{Bouchaud_1990,Schlaich_2025} and references therein). Here, we sidestep this complexity (leaving it to speculation in Section \ref{Section:4} and future work) and consider the slightly unrealistic case in which the media may be multi-scale along the magnetic-field line but has some typical correlation length~$l_{\perp}$ across it.

In such a set-up, a CR with a perpendicular diffusion coefficient~$\kappa_{\perp}$ will typically sample an entirely independent medium on a time scale~$\tau_{\perp} \sim l_{\perp}^{2}/\kappa_{\perp}$, over which time it will have travelled a parallel distance~$\sim \sqrt{\tau_{\perp} \bar{\kappa}(\tau_{\perp})}$, where $\bar{\kappa}$ is the running diffusion coefficient the medium would have had were the transport purely 1D. Since the CR samples a new medium after this time, its trajectory has effectively been reset, so it now diffuses with steps~$\sim \sqrt{\tau_{\perp} \bar{\kappa}(\tau_{\perp})}$ over times~$\tau_{\perp} \sim l_{\perp}^{2}/\kappa_{\perp}$, giving a 2D diffusion coefficient that should scale~as
\begin{equation}
\label{eqn:4:3p5}
\bar{\kappa}_{\mathrm{2D}} \sim \bar{\kappa}(t \sim l_{\perp}^{2}/\kappa_{\perp}).
\end{equation}
For a two-phase medium with pdfs of high- and low-diffusion patches as in Section~\ref{Section:2}, we find, using (\ref{eqn:2:2}),
\begin{equation}
\label{eqn:4:4}
\bar{\kappa}_{\mathrm{2D}} \sim (1-f)\kappa_{\mathrm{high}}\begin{cases} 1 - \left(\displaystyle\frac{\kappa_{\mathrm{high}}l_{\perp}^{2}}{\kappa_{\perp}l_{\mathrm{max}}^{2}} \right)^{(2-\alpha)/2} \,\, &\text{for} \,\,\, \alpha < 2, \\[15pt] \left.\ln\left(\displaystyle\frac{\kappa_{\perp}l_{\mathrm{max}}^{2}}{\kappa_{\mathrm{high}}l_{\perp}^{2}} \right)\right/ \ln\left(\displaystyle\frac{l_{\mathrm{max}}^{2}}{l_{\mathrm{min}}^{2}} \right) \,\, &\text{for} \,\,\, \alpha = 2, \\[15pt] \left(\displaystyle\frac{\kappa_{\mathrm{high}}l_{\perp}^{2}}{\kappa_{\perp} l_{\mathrm{min}}^{2}} \right)^{(2-\alpha)/2}\,\, &\text{for} \,\,\, 4 \geq \alpha > 2, \\[15pt]
\left(\displaystyle\frac{\kappa_{\mathrm{high}}l_{\perp}^{2}}{\kappa_{\perp}l_{\mathrm{min}}^{2}} \right)^{-1}\,\,\, &\text{for} \,\, \alpha > 4, 
\end{cases}
\end{equation}
with the further caveat that, for perpendicular-diffusion times~$\tau_{\perp}\ll l_{\mathrm{min}}^{2}/\kappa_{\mathrm{high}}$, one must have~$\bar{\kappa}_{\mathrm{2D}} = \kappa_{\mathrm{A}}$ (a result that it is possible to prove rigorously: see Appendix \ref{Section:A1}). 

The agreement of this result with numerical simulation Figure~\ref{Fig:C1} (see also Figure~\ref{Fig:5}) is modest compared with the success shown in Figure~\ref{Fig:3} for 1D diffusion. Nevertheless, what this result captures, manifestly, is a way in which the patchiness of the medium, through~$\alpha$, may affect the energy dependence of the CR diffusion coefficient at all times. In particular, should one desire to have a particular energy-dependent diffusion coefficient (for instance,~$E^{0.5}$ associated with galactic CR propagation, see \citealt{Evoli_2008,Trotta_2011,Hopkins_2022c}), then there may exist a particular~$\alpha$ for which an energy-dependent parallel diffusion coefficient~$\kappa_{\mathrm{low}} \sim E^{\beta}$ will give such an effective diffusion coefficient. When perpendicular diffusion is present (not so much as to overwhelm parallel transport) the stochastic inhomogeneity of the medium can imprint itself on the large-scale diffusive behaviour.
\section{Conclusion}
\label{Section:4}
There is a wealth of evidence, both numerical and observational (e.g., \citealt{Beattie_2025,Weber_2025}) that the medium through which CRs propagate is stochastically inhomogeneous. Furthermore, theoretical and numerical insights (e.g., \citealt{Armillotta_2021,Armillotta_2022,Kempski_2022,Hopkins_2022,Butsky_2024,Armillotta_2025}) have provided convincing evidence that a multi-phase medium is not just realistic, but necessary to understand CR transport in astrophysically relevant contexts. Equally, theoretical and numerical tools for understanding turbulence and intermittency of MHD-adjacent systems, have become increasingly sophisticated (see, e.g., \citealt{Schekochihin_2022} and references therein): the pdfs and exponents of whatever quantity the reader desires can be theorised or, failing that, measured numerically. 

The problem of understanding transport in a stochastically inhomogeneous media is one of identifying the minimal set of fluid quantities that describe that transport. In this paper, we have presented a toy model of such media for which the relevant quantities are especially minimal. This toy model is based on the simple, but flexible, picture of multi-scale patches of varying diffusivity threaded by magnetic-field lines (see Figure \ref{Fig:2}). In the case where transport is dominated by diffusion along magnetic-field lines, we have proven the exact result~(\ref{eqn:1:3}) that a CR diffusing with inhomogeneous diffusion coefficient $\kappa(x)$ will, over large distances, have an effective diffusion coefficient equal to the harmonic mean of the local diffusion coefficient (see Figure \ref{Fig:1}). This is the statement that, at late times, CR transport is dominated by the areas of low diffusion even if these areas occupy a relatively small fraction $f$ of the volume. In contrast, at early times, all transport is dominated by the areas of high diffusion, meaning that there is a transient period of sub-diffusivity at intermediate times and scales, which we show in~(\ref{eqn:2:2}) to depend on the power-law index~$\alpha = -\mathrm{d}\ln P/\mathrm{d}\ln l$ of the patch-size distribution~$P(l)$. A large value of $\alpha$ corresponds to a distribution dominated by very small patches, creating an abrupt jump from high to low diffusion coefficient, while a small value of $\alpha$ smooths this transition as CRs experience long uninterrupted walks through high-diffusion zones (see Figure \ref{Fig:3}). The shorthand for thinking about transport here is that over sufficiently short distances along the mean field, there will be field lines with high transport acting as highways. Over larger distances, every field line will eventually have a traffic jam, which slows transport to a crawl (cf. `Gaussianisation' considered by \citealt{Liang_2025}).

After showing that stochastically inhomogeneous media can be described, adequately for the purposes of CR transport, with a relatively small collection of parameters (filling fraction~$f$ and index~$\alpha$), we turn to the question of observables in CR transport for these media, in particular the energy dependence of the CR escape rate. While this can create breaks in the energy dependence of the CR escape rate on intermediate scales (shown in Figure \ref{Fig:4}), over large scales, the energy dependence of the escape rate is inherited directly from the underlying low-diffusion patches (the system is precisely the sum of its parts). When a small amount of perpendicular diffusion is included, we have argued that this result must change as CRs are able to avoid diffusively the regions of low diffusion (cf. navigation around large-scale magnetic mirrors in~\citealt{Hu_2025b}). This is not an expression of CRs exerting free will, but rather the statement that when perpendicular diffusion $\kappa_{\perp}$ is allowed, the route through a patchy medium is recalibrated after a typical time $l_{\perp}^{2}/\kappa_{\perp}$, giving CRs another chance to be in a high-diffusion patch. As a result, the effective parallel-diffusion coefficient (\ref{eqn:4:4}) contained both perpendicular- and parallel-diffusion coefficients in a combination controlled by $\alpha$. Since the perpendicular and parallel diffusion coefficients can have unique energy dependence, this allows for the stochastic inhomogeneity (parametrised by $\alpha$) of the medium to imprint itself on the energy dependence of CR transport (see Figure~\ref{Fig:5}). The cynical outlook of this result is that any transport can be achieved by any mechanism provided the patch-size distribution has the right scaling (a fact that is manifestly clear from (\ref{eqn:4:4})). However, the optimistic viewpoint is that multi-phase media have been reduced to a handful of parameters, the measurement of which, both in simulation and observation, is not a fruitless endeavour.

The model that we have proposed is overly simplified---deliberately so. The most obvious omission is of phases in which the CR transport is not perfectly diffusive. The most immediate of such situations are those in which the CR transport is ballistic (or streaming-dominated), but more generally, one can imagine any degree of anomalous fractional diffusivity (see, e.g., \citealt{Liang_2025,Hu_2025b}). This has been considered in great depth in a variety of other formalisms, most notably that of treating the transport as being dominated by ballistic propagation interspersed with strong-scattering events (\citealt{Kempski_2023,Lemoine_2023,Butsky_2024,Kempski_2025}). Within our formalism, one may prefer to view such alternative transport models as the micro-physical mechanism that provides macroscopic diffusion in a yet larger-scale stochastic inhomogeneity (i.e., `it's patches all the way down'). Alternatively, should there be no scale separation between micro-physical transport and stochastic inhomogeneity (i.e., if the transport is anomalous on macroscopic scales), we present in Appendix \ref{Section:D} a natural (heuristic) discussion of how parallel streaming may be incorporated into our formalism of multi-phase media. While this would have strong implications for the early-time transport (and therefore the possible energy dependence in the presence of moderate perpendicular diffusion), we argue that, provided at least one of the composite media is strongly diffusive, the long-term diffusivity will still approach the harmonic mean.

Another omission is that of neglecting the underlying fluid motion of the medium itself. Should the diffusion coefficient in the low-diffusion regions become too small, the question ought not to be one of how quickly CRs move through the medium but how quickly the medium itself is capable of moving CRs. This is naturally intertwined with the question of coupling and feedback between CRs and the fluid through which they propagate (see, e.g., \citealt{Butsky_2018,Butsky_2020, Weber_2025,Sampson_2025}). One may expect that, for sufficiently strong coupling, the turbulent diffusivity of the fluid sets the lower bound on the CR diffusion coefficient, while in the more weakly coupled case, the slippage between the CRs and the fluid may lead one to speculate that a more appropriate model for the motion of the background fluid would be a diffusivity $\kappa(x,t)$ and mean flow $u(x,t)$ that are stochastic in both space and time. \footnote{In the condensed-matter literature, this distinction is referred to as the system being `annealed' (stochastic in time) versus `quenched' (stochastic in space, constant in time), leading to distinct behaviours (see, e.g., \citealt{Bouchaud_1990,Gueneau_2025})}

Perhaps the most interesting, if slightly academic, question is the one that we have failed to address fully: in a three-dimensional diffusive medium that is multi-scale, both parallel and perpendicular to the magnetic field, how do CRs propagate? This question draws very close, as we have pointed out several times in this paper, to the issue of diffusion of particles in percolation clusters (affectionately referred to as `ants in labyrinths' by \citealt{Gennes_1976}). In condensed-matter physics, exact solutions to such systems---ones in which particles may find optimal paths, or waste long times in `culs de sac'---typically exploit simplified geometries (e.g., \citealt{Meerson_2019}), or some self-similarity (e.g., \citealt{Havlin_2002}) of the system that, given the possibly turbulent and self-similar origin of astrophysical multi-phase systems, may also be applicable for CR transport. We leave such exciting interdisciplinary considerations to future work.
\section*{Acknowledgements}
It is a pleasure to thank Thomas Foster, Phil Hopkins, Philipp Kempski, Martin Lemoine, Luca Orusa, Sid Parameswaran, Sasha Philippov, Eliot Quataert, Amrita Sahu, Anatoly Spitkovsky, and Jono Squire for their insightful comments. RJE was supported by the Simons Foundation grant MP-SCMPS-00001470; PR by a Gateway Fellowship and a Walter Benjamin Fellowship; MLN by a Clarendon Scholarship; FM by a Leverhulme Trust International Professorship grant (Award Number: LIP-2020-014); AFAB by a UKRI Future Leaders Fellowship (grant number MR/W006723/1); MWK in part by NSF CAREER Award No.~1944972. The work of AAS was supported in part by grants from STFC (ST/W000903/1) and EPSRC (EP/R034737/1), as well as by the Simons Foundation via a Simons Investigator award. This research was also supported in part by the NSF grant PHY-2309135 to the Kavli Institute for Theoretical Physics (KITP) and benefited from many vigorous interactions with the participants of their programme on ``Interconnections between the Physics of Plasmas and Self-gravitating Systems''. This work was also facilitated by the Multimessenger Plasma Physics Center (MPPC), NSF grant PHY-2206607.
\section*{Data availability}
The data underlying this article will be shared on reasonable request to the corresponding author.


\bibliographystyle{mnras}
\bibliography{CRS} 




\appendix
\section{Early time diffusion}
\label{Section:Z}
In this appendix, we prove (the somewhat obvious statement) that an ensemble of particles that diffuse according to (\ref{eqn:1:1}) will have a mean diffusion coefficient equal to the arithmetic mean of the local diffusion coefficient at early times. To do this, we consider our Green's function at early times. By definition,
\begin{equation}
n(x,t=0;x_{0}) = \delta(x-x_{0}).
\end{equation}
Then, using (\ref{eqn:1:1}) together with (\ref{eqn:1:2}), we have
\begin{equation}
\begin{split}
\left.\dev{(t \bar{\kappa}(t))}{t}\right|_{t=0} & = \left\langle\frac{1}{2}\int\dd{x}(x-x_{0})^2 \pdev{}{x}\left[\kappa(x)\pdev{n}{x} \right]_{t=0}\right\rangle_{x_{0}} \\ & =  \left\langle \int\dd{x} \left[\kappa(x) + \left(x - x_{0} \right)\pdev{\kappa}{x}\right]\delta(x-x_{0}) \right\rangle_{x_{0}} \\ & = \langle \kappa(x_{0})\rangle_{x_{0}}  = \lim_{X\to \infty}\frac{1}{X}\int\dd{X}\kappa(X) = \kappa_{\mathrm{A}}.
\end{split}
\end{equation}
Integrating this for early times gives the required result (\ref{eqn:1:2p5}):
\begin{equation}
\bar{\kappa}(t) = \kappa_{\mathrm{A}} + O\left(t\right).
\end{equation}
\section{Homogenisation in a medium with spatially dependent diffusion coefficient}
\label{Section:A}
In this appendix, we present the formal derivation of the result that, on large scales, the effective diffusion coefficient of a CR in an inhomogeneous medium in one spatial dimension is the harmonic mean of the local diffusion coefficient. 

We begin with the inhomogeneous diffusion equation
\begin{equation}
\label{eqn:A1}
\pdev{n}{t} = \pdev{}{x}\kappa(x) \pdev{n}{x},
\end{equation}
and our goal is to derive from it a diffusion equation with a large-scale homogeneous diffusion coefficient $\bar{\kappa}$, which will be the limit~{${\bar{\kappa}(t\to \infty)}$}. Physically, we assume that~$\kappa(x)$ only varies on scales up to some~$l_{\mathrm{max}}$, so we are interested in capturing the diffusion of density on scales~{${l \geq l_{\mathrm{max}}}$}. Mechanically letting~$\varepsilon = l_{\mathrm{max}}/l \ll 1$, this amounts to introducing a scaled variable~{${X = x/\varepsilon}$} that describes the small-scale variation of~$\kappa$, and thus upgrading the density to be a function~$n(x,X)$ of both the large-scale position~$x$ and the small-scale variation~$X$. In these new variables, the inhomogeneous diffusion equation~(\ref{eqn:A1}) becomes
\begin{equation}
\label{eqn:A2}
\varepsilon^{2}\pdev{n}{t} = \left(\varepsilon\pdev{}{x} + \pdev{}{X}\right)\kappa(X)\left(\varepsilon\pdev{}{x} + \pdev{}{X}\right)n.
\end{equation}
We now expand
\begin{equation}
\label{eqn:A3}
n(x,X,t) = n^{(0)}(x,X,t) + \varepsilon n^{(1)}(x,X,t) + \varepsilon^{2} n^{(2)}(x,X,t) + ...
\end{equation}
Collecting terms of order $\varepsilon^{0}$ in (\ref{eqn:A2}), we get
\begin{equation}
\label{eqn:A4}
0=\pdev{}{X}\kappa(X)\pdev{n^{(0)}}{X},
\end{equation}
which implies that $n^{(0)}(x,X) = n^{(0)}(x)$, provided $n^{(0)}$ is required to be finite for all $X$. This simply says that the lowest-order contribution to the density will be independent of the local fluctuations in the diffusion coefficient, i.e., that small-scale fluctuations are ironed out on short time scales.

At the first order in $\varepsilon$, we find 
\begin{equation}
\label{eqn:A5}
0 = \pdev{}{X}\kappa(X)\left(\pdev{n^{(0)}}{x} + \pdev{n^{(1)}}{X} \right),
\end{equation}
which is solved by
\begin{equation}
\label{eqn:A6}
n^{(1)}(x,X,t) = - X\pdev{n^{(0)}}{x} + C_{1}(x)\int_{0}^{X}\frac{\mathrm{d}X'}{\kappa(X')},
\end{equation}
where $C_{1}(x)$ is a constant (with respect to $X$) of integration. We should immediately note that this solution can grow arbitrarily large at large~$X$. This, however, would be unphysical because $X$ represents small-scale fluctuations and so the perturbation to density~$n^{(1)}(x,X)$ should be finite for all values of $X$. The only way to resolve this discrepancy is to impose the solvability condition that the perturbation expansion (\ref{eqn:A3}) should not be broken:  
\begin{equation}
\label{eqn:A7}
C_{1}(x) = \lim_{X\to \infty} \frac{X}{\displaystyle\int_{0}^{X}\!\frac{\mathrm{d}X'}{\kappa(X')}}\pdev{n^{(0)}}{x}.
\end{equation}
Physically this says that large-scale density gradients (in $x$) source small-scale density fluctuations.

Finally, at second order in $\varepsilon$, we have
\begin{equation}
\label{eqn:A8}
\begin{split}
\pdev{n^{(0)}}{t} & = \pdev{}{x}\underbrace{\kappa(X)\left(\pdev{n^{(0)}}{x} + \pdev{n^{(1)}}{X} \right)}_{=C_{1}(x)}\\& + \pdev{}{X}\kappa(X)\left(\pdev{n^{(1)}}{x} + \pdev{n^{(2)}}{X} \right).
\end{split}
\end{equation}
Integrating this equation once in $X$, we find
\begin{equation}
\label{eqn:A9}
X\left(\pdev{n^{(0)}}{t} - \pdev{C_{1}}{x}\right) = \kappa(X)\left(\pdev{n^{(1)}}{x} + \pdev{n^{(2)}}{X} \right) + C_{2}(x),
\end{equation}
where $C_{2}(x)$ is another integration constant with respect to $X$. Without further integration, it is easy to see that the right-hand side grows unbounded with $X$ unless it vanishes at $X\to \infty$. It can be made to do so by an appropriate choice of $C_{2}(x)$. With this choice, in the limit~$X\to \infty$, (\ref{eqn:A9}) becomes 
\begin{equation}
\label{eqn:A10}
\pdev{n^{(0)}}{t} = \pdev{C_{1}}{x} = \bar{\kappa}\pdevn{n^{(0)}}{x}{2},
\end{equation}
where
\begin{equation}
\label{eqn:A11}
\bar{\kappa} = \lim_{X \to \infty}\frac{X}{\displaystyle\int_{0}^{X}\frac{\mathrm{d}X'}{\kappa(X')}}
\end{equation}
is the effective diffusion coefficient on large scales. This is a mathematical expression of the intuitive statement that the locations along the magnetic-field line where the diffusion coefficient is most strongly suppressed dominate the CR transport.

The final statement, that~$\bar{\kappa}$ defined by~(\ref{eqn:A11}) is equivalent to (\ref{eqn:1:2}) evaluated at late times (if not obvious already), follows from the solution for the Green's function~$n(x,t;x_{0})$, which, by the above argument, is
\begin{equation}
\label{eqn:A12}
n^{(0)}(x,t;x_{0}) = \frac{1}{\sqrt{4\pi \bar{\kappa}t}}e^{-(x-x_{0})^{2}/4\bar{\kappa}t}.
\end{equation}
Thus, $\bar{\kappa}(t \gg l_{\mathrm{max}}^{2}/\kappa_{\mathrm{low}}) = \bar{\kappa}$.
\subsection{Case of strong perpendicular diffusion}
\label{Section:A1}
The general 2D diffusion equation,
\begin{equation}
\pdev{n}{t} = \pdev{}{x}\kappa_{\parallel}(x,y)\pdev{n}{x} + \pdev{}{y}\kappa_{\perp}(x,y)\pdev{n}{y}, 
\end{equation} 
where $x$ and $y$ represent the coordinates parallel and perpendicular to the magnetic-field line, respectively, cannot (to the authors' knowledge) be solved by the same means as above. However, in the limit where the perpendicular diffusion coefficient is large,~$\kappa_{\mathrm{\perp}}/l_{\perp}^{2} \gg \kappa_{\parallel}/l_{\parallel}^{2}$, it is possible to show that the mean parallel diffusion coefficient of CRs will be equal to the arithmetic mean of the parallel diffusion coefficient. Physically this is transparent: the CRs traverse the direction perpendicular to the magnetic-field line so quickly that their diffusivity is rapidly averaged.

Analytically, we see this by performing the same trick as above, introducing a multiple-scales variable (this time in $y$), $Y = y/\varepsilon_{\perp}$, with the expansion parameter $\varepsilon_{\perp} = l_{\mathrm{max}}/l_{\perp}$. Again we arrive at the equation
\begin{equation}
\begin{split}
\varepsilon_{\perp}^{2}\pdev{n}{t} & = \varepsilon_{\perp}^{2}\pdev{}{x}\kappa_{\parallel}(x,Y)\pdev{n}{x} \\ & + \left(\varepsilon_{\perp}\pdev{}{y} + \pdev{}{Y}\right)\kappa_{\perp}(x,y)\left(\varepsilon_{\perp}\pdev{}{y} + \pdev{}{Y}\right)n.
\end{split}
\end{equation}
Performing an expansion of~$n$ similar to~(\ref{eqn:A3}), now in the perpendicular direction, we find the solvability condition
\begin{equation}
\int\dd{Y}\left[\pdev{n^{(0)}}{t} -  \pdev{C_{1,Y}}{y}  - \pdev{}{x}\kappa_{\parallel}(x,Y)\pdev{n^{(0)}}{x} \right] = 0,
\end{equation}
where 
\begin{equation}
C_{1,Y}(x,y) = \lim_{Y\to\infty}\frac{Y}{\int_{0}^{Y}\displaystyle\frac{\mathrm{d}Y'}{\kappa_{\perp}(x,Y')}}\pdev{n^{(0)}}{y} \equiv \bar{\kappa}_{\perp}(x)\pdev{n^{(0)}}{y}.
\end{equation}

Thus, in this somewhat contrived limit of immensely fast perpendicular transport, the perpendicular diffusion coefficient takes the value of its harmonic mean, whereas the parallel diffusion coefficient takes the value of its arithmetic mean.
\section{Transient sub-diffusive transport}
\label{Section:B}
\begin{figure}
\centering
\includegraphics[width=8.68cm,height=7.62cm]{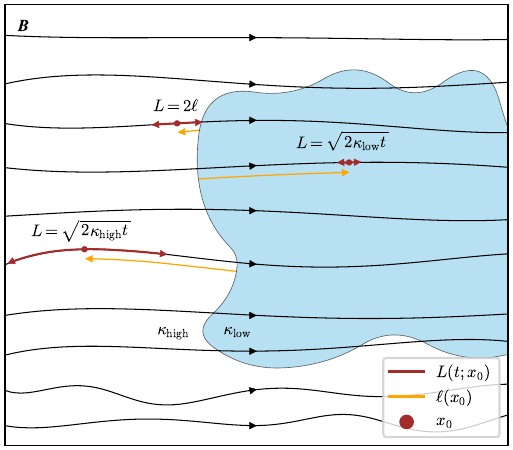}
\caption{\label{Fig:5p5}Cartoon illustrating the necessary elements for the calculation of the progressive harmonic mean~$\bar{\kappa}_{\mathrm{prog}}(t)$ given by (\ref{eqn:B2}). A CR sourced at position $x_{0}$ diffuses a distance~$L(t;x_{0})$ in time $t$, found by inverting~(\ref{eqn:B1}). In a two-phase medium, this amounts to a CR sourced in the high-diffusion region travelling unimpeded a distance~$L\sim \sqrt{2\kappa_{\mathrm{hight}}t}$, or a distance~$\ell(x_{0})$ and getting stuck in a low-diffusion region, whichever happens sooner.}
\end{figure}
While the picture of a two-phase medium illustrated in Figure \ref{Fig:2} is a useful heuristic for understanding the sub-diffusive behaviour of CRs, it clearly does not encompass an arbitrarily varying diffusion coefficient or indeed the case with more than two phases. In this appendix, we propose a (non-rigorous) scheme for computing the sub-diffusive behaviour for arbitrary~$\kappa(x)$, and show that it is consistent with Monte-Carlo simulations of the two-phase case.

The central step is again to acknowledge that the Green's function solution $n(x,t;x_{0})$ is out of reach and instead to posit, motivated by~(\ref{eqn:1:3}), that a CR starting at position~$x_{0}$ will take the time
\begin{equation}
\label{eqn:B1}
\tau(L;x_{0}) = \frac{L}{2} \int_{x_{0}-L/2}^{x_{0}+L/2}\frac{\mathrm{d}x'}{\kappa(x')},
\end{equation}
to diffuse a distance $L$ away from $x_{0}$. This is nothing more than a fitting formula (we do not expect it to be correct in any rigorous sense), but it is a well-motivated one, on the grounds that it correctly interpolates the crossing time in the two limits,~{${L\to 0}$} and~{${L \to \infty}$},~(\ref{eqn:1:2p5}) and~(\ref{eqn:1:3}), that we know analytically and that are valid for arbitrary~$\kappa(x)$. It also correctly incorporates the intuition that low-diffusion regions take more time for a CR to cross than high-diffusion ones. While the crossing time (\ref{eqn:B1}) depends on the initial CR position~$x_{0}$, it is a monotonically increasing function of~$L$, so, for each~$x_{0}$, it possesses a well defined inverse~$L(t;x_{0})$, which is the distance a CR starting at position~$x_{0}$ will have travelled in a time~$t$. From this, we can compute the mean diffusion coefficient as a function of time:
\begin{equation}
\label{eqn:B2}
\bar{\kappa}_{\mathrm{prog}}(t) = \frac{\left\langle L^{2}(t,x_{0})\right\rangle_{x_{0}}}{2t},
\end{equation}
where the average is computed over the initial positions~$x_{0}$, as in (\ref{eqn:1:2}). We refer to this running diffusion coefficient as the `progressive harmonic mean'. It naturally recovers the harmonic mean at long times, but also extends smoothly to the correct initial diffusion coefficient. We stress that this diffusion coefficient is both approximate (the assumption (\ref{eqn:B1}) is based solely on the reasonable interpolation of two known functions) and still difficult to compute analytically (being proportional to the average of a square of the inverse of a random variable). It should therefore be viewed as a numerical tool---one that agrees remarkably well with Monte-Carlo simulations of CR transport (the blue lines in Figure~\ref{Fig:3}). Nevertheless, in the spirit of practising analytically what we preach numerically, and to motivate that simple scalings like those in (\ref{eqn:2:2}) could be found for more general (multi-phase) $\kappa(x)$, we will now compute~(\ref{eqn:B2}) at early times for the case of the two-phase medium discussed in Section \ref{Section:2}.

Even in a two-phase medium, the task of inverting~$\tau(L;x_{0})$ for~$L$ to substitute into (\ref{eqn:B2}) is difficult because, for a given~$x_{0}$ and~$L$, the integration interval in~(\ref{eqn:B1}) can span multiple patches, making finding the inverse with respect to~$L$ difficult. To make progress, we will consider the limit~$\kappa_{\mathrm{low}} \ll \kappa_{\mathrm{high}}$ for times~$t < l_{\mathrm{min}}/2\kappa_{\mathrm{low}}$ (still a sizeable portion of the sub-diffusive transient in Figure~\ref{Fig:3}). In this limit, we may define~$\ell(x_{0})$ to be the distance from a given initial source of CRs to the nearest boundary between high- and low-diffusion regions (see Figure~\ref{Fig:5p5}). If~$x_{0}$ resides in a high-diffusion patch, the length~$L(t;x_{0})$ required for the travel time to equal~$t$ is either~$\sqrt{2\kappa_{\mathrm{high}}t}$ or twice the distance~$\ell(x_{0})$ to the closest boundary between a low- and high-diffusion regions, whichever is smaller (see Figure~\ref{Fig:5p5}). The result is
\begin{equation}
\label{eqn:B-3}
\frac{L^{2}(t;x_{0})}{2\kappa_{\mathrm{high}}t} = \begin{cases}
\displaystyle\frac{[2\ell(x_{0})]^{2}}{2\kappa_{\mathrm{high}}t} + O\left(\frac{\kappa_{\mathrm{low}}}{\kappa_{\mathrm{high}}}\right) \quad &\text{for}\quad 2\ell(x_{0}) \leq \sqrt{2\kappa_{\mathrm{high}}t},\\
1 \quad &\text{for} \quad 2\ell(x_{0}) \geq \sqrt{2\kappa_{\mathrm{high}}t},
\end{cases}
\end{equation}
while for positions $x_{0}$ in low-diffusion patches, only a minuscule distance is required for the travel time to equal $t$:
\begin{equation}
\label{eqn:B-2}
\frac{L^{2}(t;x_{0})}{2\kappa_{\mathrm{high}}t} = O\left(\frac{\kappa_{\mathrm{low}}}{\kappa_{\mathrm{high}}}\right).
\end{equation}
The physical picture here is that illustrated in Figure \ref{Fig:5p5}: the CRs sourced at~$x_{0}$ inside low-diffusion patches diffuse a very short distance over a time~$t$, while the CRs sourced at~$x_{0}$ close to the boundary between high- and low-diffusion regions diffuse until they contact that boundary. 

To compute the average over $x_{0}$ in (\ref{eqn:B2}), we need only know how far away a given initial source of CRs is from a boundary (given by~$\ell(x_{0})$). For a pdf of high-diffusion patches given by
\begin{equation}
\label{eqn:B-1p5}
P_{\mathrm{high}}(l) = (\alpha - 1)\frac{l^{-\alpha}}{l_{\mathrm{min}}^{1-\alpha} - l_{\mathrm{max}}^{1-\alpha}},
\end{equation}
the probability density that a given~$x_{0}$ is a distance~$\ell(x_{0})$ from a boundary between the low- and high-diffusion regions is
\begin{equation}
\label{eqn:B-1}
\mathcal{F}(\ell) = \begin{cases}
2\displaystyle\frac{\alpha-2}{\alpha-1}\displaystyle\frac{l_{\mathrm{min}}^{1-\alpha} - l_{\mathrm{max}}^{1-\alpha}}{l_{\mathrm{min}}^{2-\alpha} - l_{\mathrm{max}}^{2-\alpha}}\quad &\text{for} \quad 2\ell < l_{\mathrm{min}}, \vspace{4pt}\\ 2\displaystyle\frac{\alpha-2}{\alpha-1}\displaystyle\frac{\left(2\ell\right)^{1-\alpha} - l_{\mathrm{max}}^{1-\alpha}}{l_{\mathrm{min}}^{2-\alpha} - l_{\mathrm{max}}^{2-\alpha}} \quad &\text{for} \quad 2\ell \geq l_{\mathrm{min}}.
\end{cases}
\end{equation}
The average~(\ref{eqn:B2}) is then just an integral over all~$x_{0}$, equivalent to an integral over all~$\ell$, weighted by $\mathcal{F}$: 
\begin{equation}
\label{eqn:B-0}
\bar{\kappa}_{\mathrm{prog}}(t) = \frac{1-f}{2t}\displaystyle\int_{0}^{l_{\mathrm{max}}/2}\!\!\!\!\!\!\!\!\!\!\dd{\ell}\mathcal{F}(\ell)L^{2}(t;x_{0}).
\end{equation}

Using the combination of~(\ref{eqn:B-1}) and~(\ref{eqn:B-3}), it is possible (with significant stamina), to check that, to leading order in~$\kappa_{\mathrm{low}}/\kappa_{\mathrm{high}}$,~(\ref{eqn:B-0}) agrees with our expressions~(\ref{eqn:2:1}) and (\ref{eqn:2:2}), derived on heuristic grounds in Section~\ref{Section:2}. At very early times~$t < l_{\mathrm{min}}^{2}/2\kappa_{\mathrm{high}}$, we have
\begin{equation}
\bar{\kappa}_{\mathrm{prog}}(t) = (1-f)\kappa_{\mathrm{high}}\left(1 - \frac{2}{3}\frac{\sqrt{2\kappa_{\mathrm{high}}t}}{\crl{l}} \right),
\end{equation}
where~$\crl{l}$ is the mean patch length associated with~(\ref{eqn:B-1p5}). At very late times~$t > l_{\mathrm{max}}^{2}/2\kappa_{\mathrm{high}}$, one finds
\begin{equation}
\bar{\kappa}_{\mathrm{prog}}(t) = \frac{1-f}{2\crl{l}t}\left(l_{\mathrm{min}}^{3} + \crl{l^{3}} - l_{\mathrm{max}}^{1-\alpha}\crl{l^{\alpha+2}} \right).
\end{equation}
For intermediate times~$l_{\mathrm{min}}^{2}< 2\kappa_{\mathrm{high}}t < l_{\mathrm{max}}^{2}$, (\ref{eqn:B-0}) leads to a proliferation of terms that are not particularly illuminating. To demonstrate that the result is nevertheless correct, consider the case~$2<\alpha < 4$ in the limit of~$l_{\mathrm{max}} \gg l_{\mathrm{min}}$. This gives, after some algebra,
\begin{equation}
\begin{split}
\bar{\kappa}_{\mathrm{prog}}(t)  =& \frac{1-f}{\alpha - 1}\kappa_{\mathrm{high}}\left(\frac{2\kappa_{\mathrm{high}}t}{l_{\mathrm{min}}^{2}} \right)^{(2-\alpha)/2} + \frac{1-f}{6t}l_{\mathrm{min}}^{2}
\\ &+ \frac{1-f}{2t l_{\mathrm{min}}^{2-\alpha}}\frac{\alpha-2}{(\alpha-1)(4-\alpha)}\left[\left( 2\kappa_{\mathrm{high}}t\right)^{(4-\alpha)/2} - l_{\mathrm{min}}^{4-\alpha} \right],\end{split}
\end{equation} 
which recovers with the scalings~(\ref{eqn:2:2}).

Thus, the heuristic that the progressive harmonic mean~$\bar{\kappa}_{\mathrm{prog}}(t)$ is approximately equal to the true running diffusion coefficient~$\bar{\kappa}(t)$ is valid. This heuristic has the advantage that no CRs actually need to be propagated through the system to compute~$\bar{\kappa}_{\mathrm{prog}}(t)$, and it may be useful in computing diffusion coefficients with arbitrarily varying~$\kappa(x)$. The central feature, however, is that it gives rise to strong quantitative agreement in Figure~\ref{Fig:3}, which provides hope for analytical progress in future work.
\section{Numerical methods}
\label{Section:C}
\begin{figure*}
\centering
\includegraphics[width=17.78cm,height=10.67cm]{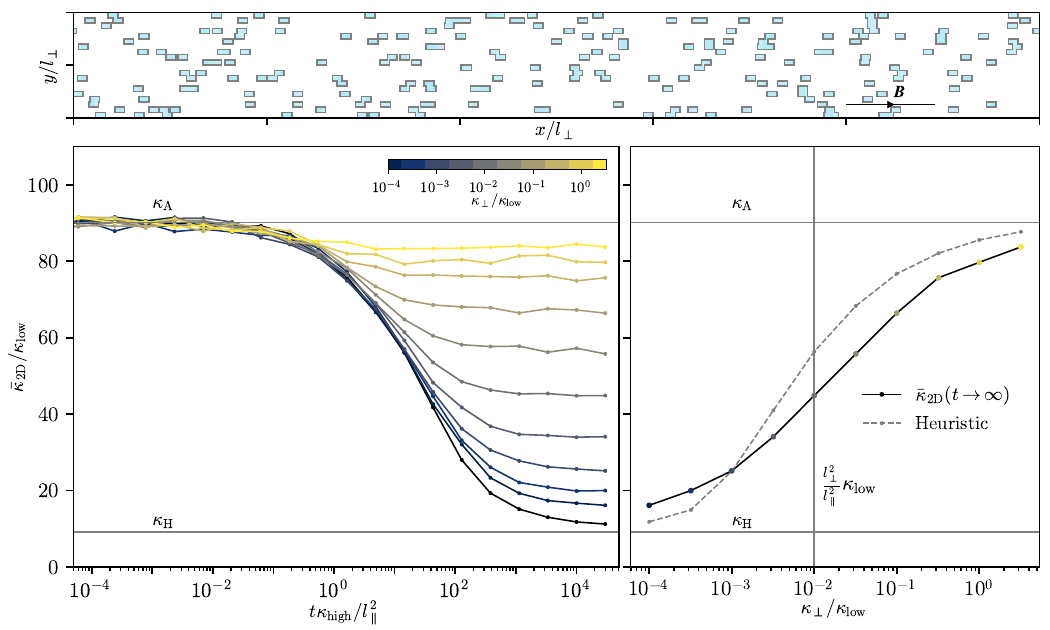}
\caption{\label{Fig:5}Same as Figure \ref{Fig:C1} for a 2D diffusive medium with single-scale low diffusion patches ($l_{\parallel}/l_{\perp} = 10$ and $f = 0.1$)  and $\kappa_{\mathrm{high}}/\kappa_{\mathrm{low}} = 10^{2}$. Again the converged values are in reasonable agreement with the heuristic~(\ref{eqn:4:3p5}).}
\end{figure*}
In this appendix, we detail our numerical solution to the CR diffusion problem. This is simply a Monte-Carlo diffusion problem across 1D and 2D patchy diffusive environments. Each CR has a continuous spatial coordinate~$x(t)$ for which discrete time steps~$\Delta t$ are made with the rule
\begin{equation}
\label{eqn:C:1}
x(t+ \Delta t) = x(t) + \sqrt{2\kappa(x(t))\Delta t}\chi(t) + \left.\pdev{\kappa}{x}\right|_{x = x(t)}\!\!\!\!\!\!\!\!\!\!\Delta t,
\end{equation}
where, $\chi(t)$ is a normally distributed, unit-variance random number that is independent at each time step, i.e., $\crl{\chi(t)} = 0$, and~$\crl{\chi(t)\chi(t')} = \delta_{t,t'}$. When the diffusion was in 2D, as for Figure~\ref{Fig:C1}, the random walk in~$x$ was still inhomogeneous and given by~(\ref{eqn:C:1}), while the random walk in $y$ had a constant diffusivity $\kappa_{\perp}$, viz,
\begin{equation}
y(t+\Delta t) = y(t) + \sqrt{2\kappa_{\perp}\Delta t}\chi(t).
\end{equation}

The inhomogeneous medium for 1D diffusion was generated by constructing pdfs of high-diffusion patch sizes~$P_{\mathrm{high}}(l)$ and low-diffusion patch sizes~$P_{\mathrm{low}}(l)$, and stacking samples of these pdfs `back to back' (see Figures~\ref{Fig:2} and~\ref{Fig:1}) to create an alternating sequence of high- and low-diffusion patches. This ignored any possible correlations between patches of different lengths (the answer would, potentially, be different if high- and low-diffusion patches came in pairs of correlated size; we do not consider this feature in our model). For 2D diffusive environments, we stack these 1D environments vertically (this method currently does not allow for the obvious extension to patches of varying perpendicular correlation lengths). When both the high- and low-diffusion patches are single scale (see Figure~\ref{Fig:5}) choosing pdfs of patches sizes to be delta distributed would, undesirably, produce precisely periodic media. For the simulation shown in Figure~\ref{Fig:5} therefore, we chose pdfs which were not broad (i.e., all moments in $l$ exist) and  possessed the correct mean patch lengths.

Because the time stepping~(\ref{eqn:C:1}) requires the diffusion coefficient to be differentiable, we smoothed the resulting patch-size distribution on a scale~$l_{\mathrm{min}}/10$. In order to ensure that the Monte-Carlo process successfully simulates~(\ref{eqn:1:1}), a particle obeying~(\ref{eqn:C:1}) must not (typically) jump over the smoothing scale~$l_{\mathrm{min}}/10$: we therefore required
\begin{equation}
\frac{\kappa \Delta t}{l_{\mathrm{min}}^{2}} \lesssim 10^{-2}.
\end{equation} 

Of course, no CR motion is actually diffusive on the microscopic level. At the smallest length scales, we should expect CRs to undergo field-line-following gyromotion with scattering events changing their pitch angle and gyrocenter. To confirm the (obvious, but assumed) fact that a particle undergoing rapid pitch-angle scattering also obeys the inhomogeneous diffusion equation, we supplemented the position-space random walks with random walks in pitch angle. Naturally, in this set-up, one should expect identical results on scales longer than the mean free path. In this formulation, CRs were assigned a continuous spatial coordinate $x(t)$ and pitch angle $\theta(t)$, updated according to the rules
\begin{equation}
x(t+\Delta t) = x(t) + v\Delta t\cos\theta(t)
\end{equation}
and 
\begin{equation}
\theta(t + \Delta t) = \theta(t) + \sqrt{2\sigma(x(t)) \Delta t}\chi(t),
\end{equation}
where $\chi(t)$ was, again, a normally distributed random number with unit variance. The velocity $v$ and spatially dependent scattering rate~$\sigma$ were chosen so that~$v^{2}/\sigma(x(t)) = \kappa(x)$. The running diffusion coefficients of these CRs are shown in red in Figure~\ref{Fig:3} for late times. For all simulations shown in this paper, we ensured that the mean free path $v/\sigma$ was smaller than the minimum patch size of high- and low-diffusion patches. In Appendix \ref{Section:D}, we consider how the system's behaviour would be expected to change if this condition were to be relaxed.
\section{The effect of ballistic or Alfv\'{e}nic streaming}
\label{Section:D}
\begin{figure}[t]
\centering
\includegraphics[width=8.68cm,height=7.62cm]{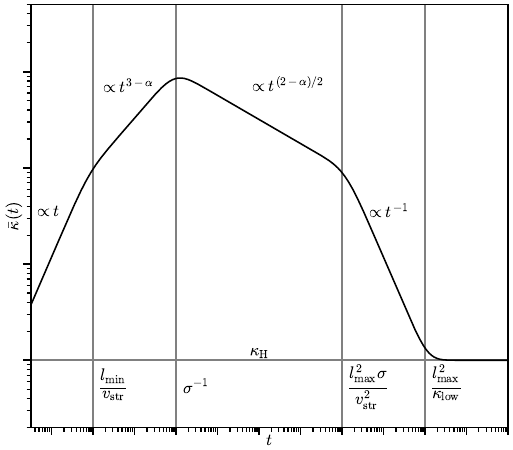}
\caption{\label{Fig:6}Cartoon of the running diffusion coefficient for a two-phase medium in which one phase is diffusive while the other allows CRs to propagate ballistically at speed~$v_{\mathrm{str}}$ with scattering rate~$\sigma$.}
\end{figure}
In this appendix, we discuss the most na\"{i}ve extension of our physical picture to the case where the CRs are not diffusive but instead stream through the medium at a specific speed $v_{\mathrm{str}}$ with a scattering rate~$\sigma$. For such CRs within this single `streaming medium', the running diffusion coefficient would follow the rough scaling
\begin{equation}
\label{eqn:D:1}
\kappa_{\mathrm{str}}(t) \sim \frac{v_{\mathrm{str}}^{2}t}{1 + \sigma t}.
\end{equation}
One could (and should) then conceptualise the diffusive media that we discuss in the main text as being realisations of this streaming medium in which the mean free path $\sim v_{\mathrm{str}}/\sigma$ is much smaller than the other length scales of interest ($l_{\mathrm{min}}$ and $l_{\mathrm{max}}$) so that the motion is diffusive on all scales of interest. Here we ask what occurs in a two-phase medium of which one phase is diffusive, with diffusion coefficient~$\kappa_{\mathrm{low}}$, while the other enables ballistic transport with a mean free path that is comparable to~($l_{\mathrm{min}} \ll v_{\mathrm{str}}/\sigma \ll l_{\mathrm{max}}$) or much greater than~($v_{\mathrm{str}}/\sigma \gg l_{\mathrm{max}}$) all scales of interest. 

In the long-time limit, the behaviour of the system will again be diffusive with approximate scaling $\bar{\kappa}(t) \to \kappa_{\mathrm{low}}/f$. At early times, we may use an argument identical to that which led to (\ref{eqn:2:1}) save for the fact that the diffusion coefficient $\kappa_{\mathrm{high}}$ is replaced by (\ref{eqn:D:1}):

\begin{equation}
\bar{\kappa}(t) \sim (1-f)\kappa_{\mathrm{str}}(t)\begin{cases} 1 - \left(\displaystyle\frac{\kappa_{\mathrm{str}}(t)t}{l_{\mathrm{max}}^{2}} \right)^{(2-\alpha)/2} \quad &\text{for} \quad \alpha < 2, \\[15pt] \left.\ln\left(\displaystyle\frac{l_{\mathrm{max}}^{2}}{t \kappa_{\mathrm{str}}(t)} \right)\right/ \ln\left(\displaystyle\frac{l_{\mathrm{max}}^{2}}{l_{\mathrm{min}}^{2}} \right) \quad &\text{for} \quad \alpha = 2, \\[15pt] \left(\displaystyle\frac{t \kappa_{\mathrm{str}}(t)}{l_{\mathrm{min}}^{2}} \right)^{(2-\alpha)/2}\quad &\text{for} \quad 4 \geq \alpha > 2, \\[15pt]
\left(\displaystyle\frac{t \kappa_{\mathrm{str}}(t)}{l_{\mathrm{min}}^{2}} \right)^{-1}\quad &\text{for} \quad \alpha > 4. 
\end{cases}
\end{equation}
We show a cartoon example of such transport in Figure \ref{Fig:6}. Despite the diversity of diffusive and sub-diffusive scalings that can arise from different values of $\alpha = -\mathrm{d}\ln P /\mathrm{d}\ln l$, it should again be stressed that, without the effect of perpendicular diffusion, the same argument as in Section \ref{Section:3} still applies: the inhomogeneity will not imprint itself on the energy dependence of the escape rate of CRs without perpendicular diffusion. 

\bsp	
\label{lastpage}
\end{document}